\documentclass[aps,reprint,prl,superscriptaddress,longbibliography,floatfix]{revtex4-2}

\usepackage{hyperref}
\usepackage{graphicx}
\usepackage{amsmath}
\usepackage{bm}
\usepackage{braket}
\usepackage{color}

\begin{document}

\title{Exchange energy of the ferromagnetic electronic ground-state in a monolayer semiconductor}

\author{Nadine Leisgang}
\affiliation{Department of Physics, University of Basel, Klingelbergstrasse 82, 4056 Basel, Switzerland}
\affiliation{Department of Physics, Harvard University, Cambridge, Massachusetts 02138, US}

\author{Dmitry Miserev}
\affiliation{Department of Physics, University of Basel, Klingelbergstrasse 82, 4056 Basel, Switzerland}

\author{Hinrich Mattiat}
\affiliation{Department of Physics, University of Basel, Klingelbergstrasse 82, 4056 Basel, Switzerland}

\author{Lukas Schneider}
\affiliation{Department of Physics, University of Basel, Klingelbergstrasse 82, 4056 Basel, Switzerland}

\author{Lukas Sponfeldner}
\affiliation{Department of Physics, University of Basel, Klingelbergstrasse 82, 4056 Basel, Switzerland}

\author{Kenji Watanabe}
\affiliation{Research Center for Electronic and Optical Materials, National Institute for Materials Science, 1-1 Namiki, Tsukuba 305-0044, Japan}

\author{Takashi Taniguchi}
\affiliation{Research Center for Materials Nanoarchitectonics, National Institute for Materials Science,  1-1 Namiki, Tsukuba 305-0044, Japan}

\author{Martino Poggio} 
\affiliation{Department of Physics, University of Basel, Klingelbergstrasse 82, 4056 Basel, Switzerland}

\author{Richard J. Warburton}
\affiliation{Department of Physics, University of Basel, Klingelbergstrasse 82, 4056 Basel, Switzerland}

\begin{abstract}
Mobile electrons in the semiconductor monolayer-MoS$_2$ form a ferromagnetic state at low temperature. The Fermi sea consists of two circles, one at the $K$-point, the other at the $\tilde{K}$-point, both with the same spin. Here, we present an optical experiment on gated MoS$_2$ at low electron-density in which excitons are injected with known spin and valley quantum numbers. The resulting trions are identified using a model which accounts for the injection process, the formation of antisymmetrized trion states, electron-hole scattering from one valley to the other, and recombination. The results are consistent with a complete spin polarization. From the splittings between different trion states, we measure the exchange energy, $\Sigma$, the energy required to flip a single spin within the ferromagnetic state, as well as the intervalley Coulomb exchange energy, $J$. We determine $\Sigma=11.2$~meV and $J=5$~meV at $n=1.5 \times 10^{12}$~cm$^{-2}$, and find that $J$ depends strongly on the electron density, $n$.
\end{abstract}

\maketitle

Ferromagnetism represents one of the canonical magnetic states. It describes a state of matter in which spontaneous alignment of electron spins leads to a net magnetization. A key metric of a ferromagnet is the exchange energy, $\Sigma$, the energy required to flip one spin. $\Sigma$ also determines the Curie temperature separating the ferromagnetic (magnetically ordered) and the paramagnetic (magnetically disordered) ground state. For the well-known metallic ferromagnets, e.g.\ iron, $\Sigma$ is large, $\sim 100$~meV, resulting in enormous Curie temperatures, $\sim 1,000$~K. The phase transition is second order and can be described by the Stoner mechanism.

Ferromagnetic ordering of mobile electrons has been observed in various two-dimensional (2D) systems, e.g., in monolayer MoS$_2$ \cite{Roch2019}, in an AlAs quantum well \cite{Hossain2021}, in monolayer WSe$_2$ \cite{Hao2022}, and in twisted bilayer graphene \cite{Sharpe2019}. As the Mermin-Wagner theorem precludes magnetic order in 2D for isotropic spins \cite{Mermin1966}, magnetic anisotropy induced by, e.g., spin-orbit interaction or a small Zeeman splitting of the Fermi surfaces is required to stabilize the ferromagnetic order of a 2D electron gas (2DEG). The zero-temperature ferromagnetic phase transition controlled by the electron density is predicted to be of the first order \cite{Miserev2019}, an idea supported experimentally \cite{Roch2020}. 

Here, we present photoluminescence (PL) with quasi-resonant excitation on gated monolayer MoS$_{2}$ in all four polarization channels. We argue that the splitting between different emission lines provides a direct measurement of the ferromagnetic exchange energy, $\Sigma$, as well as the intervalley Coulomb exchange energy, $J$. 

\begin{figure}[t!]
\centering
\includegraphics[width=85mm]{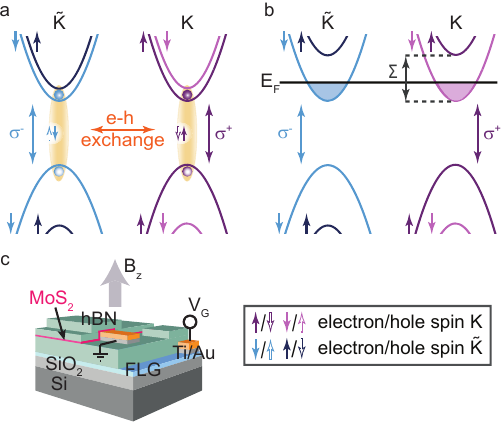}
\caption{(a) Band structure of monolayer MoS$_2$ showing exciton formation at the $K$- and $\tilde{K}$-points, and the intervalley scattering via electron-hole exchange. (b) Schematic of the reconstructed band structure containing ferromagnetically-ordered itinerant electrons with spin-$\downarrow$. (c) Schematic of the sample design. FLG stands for few-layer graphene.}
\label{bands}
\end{figure}

\begin{figure*}[t!]
\centering
\includegraphics[width=153mm]{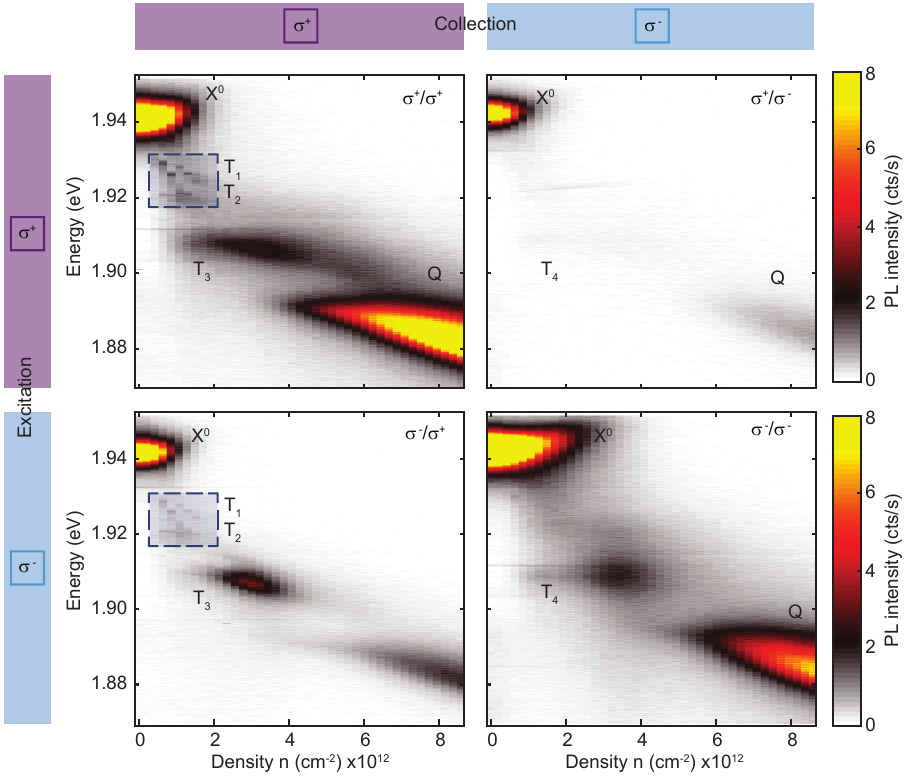}
\caption{PL for quasi-resonant excitation on gated monolayer MoS$_2$ at $+$9.00~T and 4.2~K shown as a matrix: excitation in $\sigma^{+}$ or $\sigma^{-}$, collection in $\sigma^{+}$ or $\sigma^{-}$.}
\label{PLmap}
\end{figure*}

Monolayer MoS$_2$ is a semiconductor with direct band-gaps at the $K$- and at the $\tilde{K}$-point of the Brillouin zone \cite{Mak2010}, Fig.~\ref{bands}(a). The spin-orbit splitting is large in the valence band ($\sim 150$~meV \cite{Liu2013}) and small in the conduction band (a few meV \cite{Liu2013,Kormanyos2014,Marinov2017}). Resonant $\sigma^{+}$-polarized ($\sigma^{-}$-polarized) light creates an exciton at the $K$-point ($\tilde{K}$-point). According to recent experiments, an electron gas in monolayer MoS$_2$ is ferromagnetically ordered for electron densities $n \leq 3 \times 10^{12}$~cm$^{-2}$ such that the Fermi surface consists of a circle at the $K$-point and a circle at the $\tilde{K}$-point \cite{Roch2019}. The experiments probe the electronic ground-state via the optical response: in an applied magnetic field, there is a very pronounced optical dichroism. If the spins point down, the $K \downarrow$ and $\tilde{K} \downarrow$ bands are occupied up to the Fermi energy; conversely, the $K \uparrow$ and $\tilde{K} \uparrow$ bands are pushed above the Fermi energy by the Coulomb interactions amongst the electrons and are unoccupied, as sketched in Fig.~\ref{bands}(b). The energy separation between the $\uparrow$ and $\downarrow$ bands is $\Sigma$, the exchange energy. The close-to-complete spin polarization implies that $\Sigma$ must be larger than the Fermi energy. On increasing the density, the dichroism disappears rather abruptly at a particular density, evidence of a first-order transition from a ferromagnetic state to a paramagnetic state \cite{Roch2019}. These experimental observations are consistent with theory which predicts both spin ordering (but not valley ordering) and a first-order phase transition driven by subtle corrections to Fermi-liquid theory \cite{Miserev2019}. The goal here is to determine $\Sigma$ for low electron density, $n$. 

The sample consists of a MoS$_2$ monolayer sandwiched between two hBN layers \cite{Cadiz2017,Ajayi2017}, Fig.~\ref{bands}(c). Electrons are injected into the monolayer via a gate electrode; the electron density, $n$, is proportional to the applied electrode voltage, with a capacitance calculated from the device geometry. We perform a quasi-resonant, quasi-local PL experiment: the laser photon-energy is 1.96~eV, just above the exciton energy, 1.94~eV; the PL is collected from a region with diameter 500~nm. The excitation is either $\sigma^{+}$- or  $\sigma^{-}$-polarized, thereby injecting an exciton with spin-$\uparrow$ at the $K$-point or spin-$\downarrow$ at the $\tilde{K}$-point, respectively. The PL is detected with $\sigma^{+}$- or  $\sigma^{-}$-polarization, allowing via the selection rules the responsible valley for each emission line to be determined. In PL, an electron-hole pair is injected in a particular valley, and the valley in which recombination takes place identified. In contrast, in absorption, only the eigenstates of the system are probed. A magnetic field (perpendicular to the 2D layer) of $+9.00$~T is applied: it is required to stabilize the ferromagnetic order against the Mermin-Wagner effect. The direction of the magnetic field is such that only spin-$\downarrow$ bands are occupied. The optical response is plotted as a matrix, Fig.~\ref{PLmap}: $\sigma^{+}/\sigma^{-}$ refers to excitation with $\sigma^{+}$, collection with $\sigma^{-}$; and etc.

\begin{figure*}[btp]
\centering
\includegraphics[width=180mm]{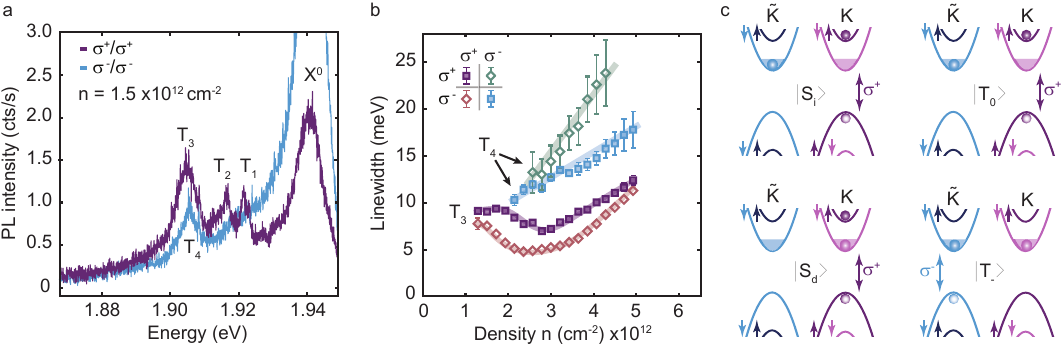}
\caption{(a) PL spectra at $n=1.5 \times 10^{12}$~cm$^{-2}$ (at $+$9.00~T and 4.2~K) for $\sigma^{+}/\sigma^{+}$ excitation/collection, and $\sigma^{-}/\sigma^{-}$ excitation/collection. (b) Trion linewidths versus $n$. (c) Schematic of the trion eigenstates showing in each case the two electron states and the hole state from which the trion is constructed.}
\label{spectra}
\end{figure*}

We focus initially on $\sigma^{+}$ excitation. At $n=0$, there is one PL line in both $\sigma^{+}/\sigma^{+}$ and $\sigma^{+}/\sigma^{-}$ corresponding to the neutral exciton, X$^{0}$. The dichroism $D=[I(\sigma^{+})-I(\sigma^{-})]/[I(\sigma^{+})+I(\sigma^{-})]$ is 42\%. On increasing $n$, X$^{0}$ weakens. In $\sigma^{+}/\sigma^{+}$, several trions are observed yet in $\sigma^{+}$/$\sigma^{-}$, the PL is very weak such that $D$ increases to $D \simeq 64$\% at
$n=1.5 \times 10^{12}$~cm$^{-2}$. 

\begin{figure}[b!]
\centering
\includegraphics[width=85mm]{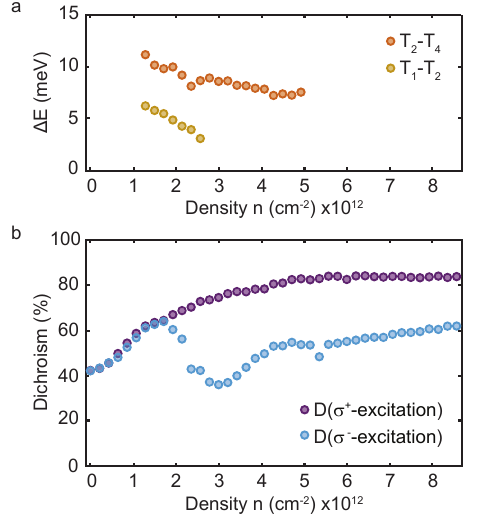}
\caption{(a) Energy splitting, $\Delta E$, versus $n$. (b) $n$-dependence of the optical dichroism, $D$, for $\sigma^+$- and $\sigma^-$-excitation.}
\label{Exchange}
\end{figure}

We propose that the $n$-dependence of $D$, Fig.~\ref{Exchange}(b), is a consequence of a large Bir-Aronov-Pikus electron-hole exchange rate. At $n=0$, an exciton injected into the $K$-valley can be scattered within its lifetime to the $\tilde{K}$-valley by the electron-hole exchange, Fig.~\ref{bands}(b). This reduces $D$ from the high value expected from the selection rules alone. Assuming an exciton lifetime of $\sim 4$~ps \cite{cadiz2016,korn2011} and that the dynamics can be described with a rate equation, the measured $D$ implies a $K \rightarrow \tilde{K}$ scattering time of $\sim 6$~ps (see Supplemental Material), consistent with experiments in the time domain \cite{mai2014}. For finite $n$, the spin-$\downarrow$ electron-states at the $\tilde{K}$-valley are occupied such that the scattering process is inhibited by the Pauli principle and $D$ increases. The increase of $D$ with $n$ is evidence that the relevant $K \leftrightarrow \tilde{K}$ scattering mechanism is electron-hole exchange, and that the $\tilde{K} \downarrow$-states become occupied.

At low $n$, three trions are observed in $\sigma^{+}/\sigma^{+}$, labelled T$_1$, T$_2$, and T$_3$, Fig.~\ref{PLmap} and Fig.~\ref{spectra}(a). T$_1$ and T$_2$ are linked: they have similar intensities and linewidths. In $\sigma^{+}/\sigma^{-}$, there is very weak PL from a trion, labelled T$_4$, Fig.~\ref{PLmap} and Fig.~\ref{spectra}(a). The energy of T$_4$ is close to that of T$_3$. However, the $n$-dependence of the T$_3$ and T$_4$ linewidths are quite different, Fig.~\ref{spectra}(b), indicating that T$_3$ and T$_4$ arise from different trion species. (We use the $n$-dependence of the trion linewidths as a diagnostic tool to identify the trions; we note that a microscopic model of the $n$-dependence is currently lacking.) 

We turn to $\sigma^{-}$-excitation. Using again the trion energies and $n$-dependent linewidths to identify the trions, in $\sigma^{-}/\sigma^{+}$ T$_1$, T$_2$, and T$_3$ are observed; in $\sigma^{-}/\sigma^{-}$ T$_4$ is observed. Hence, the collection channel and not the excitation channel determines which trions are observed. 

To proceed, it is necessary to identify the trions ${\rm T}_1 \dots {\rm T}_4$ in terms of a microscopic model (see Supplemental Material). We describe the trions in the limit of low density where the Fermi wavelength is much larger than the trion size, $\sim 2$~nm \cite{zhang2014,christopher2017,rana2020}. (At higher $n$, the eigenstates are exciton-Fermi sea polarons \cite{Suris2001,Sidler2016,Efimkin2017,Roch2019,Glazov2020}.) The low-density limit applies to the lowest $n$ used in the experiment. Electrons in MoS$_2$ have two degrees of freedom, spin $S_z = \pm \frac{1}{2}$ and valley $\tau_z = \pm \frac{1}{2}$ ($+\frac{1}{2}$ for $K$ and $-\frac{1}{2}$ for $\tilde{K}$). According to the Pauli exclusion principle, the total wave function of a trion must be antisymmetric with respect to particle exchange \cite{zhumagulov2020,Klein2022}. The two electrons within the trion bound state have therefore six eigenstates $\ket{S,S_z; \tau, \tau_z}$ characterized by the total spin $S$, its projection $S_z$, the valley pseudospin $\tau$ and its projection $\tau_z$. Four of the six trion states are relevant here:
\begin{eqnarray}
\ket{0,0; 1,1} & \equiv & \ket{S_d} \nonumber \\
\ket{0,0; 1,0} & \equiv & \ket{S_i} \nonumber \\
\ket{1,0; 0,0} & \equiv & \ket{T_0} \nonumber \\
\ket{1,-1; 0,0} & \equiv & \ket{T_-}
\label{states}
\end{eqnarray}
and are shown pictorially in Fig.~\ref{spectra}(c). $|S_d\rangle$ is the intravalley spin-singlet at the $K$-point; $|S_i\rangle$ the intervalley spin-singlet; and $|T_0\rangle$ and $|T_-\rangle$ are two spin components of the intervalley spin-triplet. 

Consider $\sigma^{+}$ excitation that creates a bright exciton at the $K$-point. The injected electron-state is $\ket{K \uparrow}$. This electron binds with a second electron to form a trion. Binding to a second $K \uparrow$ electron is impossible due to the Pauli exclusion principle. If the second electron is $K \downarrow$, the electrons form the intravalley spin-singlet state $\ket{S_d}$, see Eq.~(\ref{states}). The second spin can reside in the opposite valley, but only spin-$\downarrow$ electrons are available in the ferromagnetically ordered state. The antisymmetrized state formed is $\frac{1}{\sqrt{2}}[\ket{K_1 \uparrow_1; \tilde{K}_2 \downarrow_2} - \ket{K_2 \uparrow_2; \tilde{K}_1 \downarrow_1}]$. This state is not an eigenstate: it decomposes to $\frac{1}{\sqrt{2}} [\ket{T_0} + \ket{S_i}]$, and gives rise to two lines in the spectrum, one at the $\ket{T_0}$-energy, the other at the $\ket{S_i}$-energy. Under $\sigma^{+}/\sigma^{+}$, the lowest-energy trion T$_3$ is thereby identified as $\ket{S_d}$; the excitons in the higher-energy pair, T$_1$ and T$_2$, are identified as $\ket{S_i}$ and $\ket{T_0}$. The model explains the observation that T$_1$ and T$_2$ are linked: the lines arise from recombination of the same state.

Switching to $\sigma^{-}$-excitation, a bright exciton is created at the $\tilde{K}$-point. The injected electron-state is now $\tilde{K} \downarrow$. In the presence of only spin-$\downarrow$ electrons, the only trion that can be formed is $\ket{T_-}$. Under $\sigma^{-}/\sigma^{-}$, only T$_4$ is observed. T$_4$ is thereby identified as $\ket{T_-}$.

Finally, we analyze the cross-channels. Under $\sigma^{+}/\sigma^{-}$ conditions, the bright exciton at the $K$-point is scattered to the $\tilde{K}$-point by electron-hole exchange. Only spin-$\downarrow$ electrons are available such that the only possible trion is $\ket{T_-}$. This is consistent with the observation of T$_4$ in the spectrum. Under $\sigma^{-}/\sigma^{+}$ conditions, the bright exciton at the $\tilde{K}$-point is scattered to the $K$-point, making a spin-$\uparrow$ electron available, leading to the formation of $\ket{S_d}$ and $\frac{1}{\sqrt{2}} [\ket{T_0} + \ket{S_i}]$, such that lines T$_1$, T$_2$, and T$_3$ appear in the spectrum, exactly as observed.

The model gives a consistent description of the lines in the PL matrix and is consistent with a two-band, spin-$\downarrow$ ferromagnetism. If spin-$\uparrow$ states were occupied in the Fermi sea then a $\ket{S_d}$-like trion (specifically, $\ket{0,0; 1,-1}$) would be observed under $\sigma^{-}/\sigma^{-}$ conditions. This is not the case. Furthermore, a doublet corresponding to $\frac{1}{\sqrt{2}} [\ket{T_0} + \ket{S_i}]$ would be observed under $\sigma^{-}/\sigma^{-}$ conditions -- this is also not the case. Thus, only the spin-$\downarrow$ bands in each valley are occupied.
 
We now consider the energies of the states (see Supplemental Material), first, states $\ket{T_0}$ and $\ket{T_-}$. In a single-particle interpretation, these two states would be split by a Zeeman energy on a few-meV energy scale. (Using the spin and valley g-factors \cite{Kormanyos2014}, the single-particle splitting between $\ket{T_0}$ and $\ket{T_-}$ is $-1.03$~meV.) This is not the case: $\ket{T_0}$ and $\ket{T_-}$ are split by a much larger energy, $\simeq 10$~meV, see Fig.~\ref{Exchange}(a). The explanation lies in that $\Sigma$ contributes to $\ket{T_0}$ but not to $\ket{T_-}$. Subtracting the Zeeman splitting, we find $\Sigma \approx 11.2 \pm 1.4\,$meV at $n=1.5 \times 10^{12}$~cm$^{-2}$. The uncertainty margin arises from random noise in the PL. We note that at this density, the Fermi energy is 2.6~meV (taking the electron mass of $0.7 m_0$ \cite{Pisoni2018}), much smaller than $\Sigma$, as required for the consistency of Fig.~\ref{bands}(b). 

Second, the splitting between $|T_0\rangle$ and $|S_i\rangle$ is only possible due to an intervalley Coulomb exchange interaction, $J$, that lowers the energy of the spin-triplet $|T_0\rangle$ with respect to the spin-singlet $|S_i\rangle$, similar to Hund's rule in atoms. The splitting between T$_1 = |S_i\rangle$ and T$_2 = |T_0\rangle$ provides us with $J$ as a function of $n$, see Fig.~\ref{Exchange}(a). We extract $J \approx 5$~meV at $n = 1.5 \times 10^{12}\,$cm$^{-2}$, indicating the importance of the intervalley Coulomb exchange scattering, as pointed out in Ref.~\cite{Miserev2019}. $J$ decreases at larger $n$, see Fig.~\ref{Exchange}(a). The spin-down quantum states below the Fermi energy are occupied in the ferromagnetically ordered phase such that these states are excluded from the spin-down component of the trion state. Conversely, the spin-up quantum states remain unoccupied such that the spin-up electron component of the trion does not depend on $n$. The overlap between the spin-up and the spin-down electron densities within the trion state decreases with $n$ and tends to zero at $k_F \gg 1/a_{\rm tr}$, where $a_{\rm tr}$ is the trion size, $k_F = \sqrt{2 \pi n}$ the Fermi momentum in the ferromagnetic phase. This allows us to estimate the trion size $a_{\rm tr} \approx 1/\sqrt{\pi n_0} \approx 3$~nm, a value consistent with previous research \cite{zhang2014,christopher2017,rana2020}. Here, $n_0 \approx 3.5 \times 10^{12}$~cm$^{-2}$ is the density where $J \approx 0$~meV in Fig.~\ref{Exchange}(a).

A key component to this analysis is the observation of the T$_1 = \ket{S_i}$, T$_2 = \ket{T_0}$ ``doublet", Fig.~\ref{spectra}(a), not resolved in previous experiments \cite{Roch2019,Roch2020}. In absorption \cite{Roch2019}, two peaks were observed in $\sigma^{+}$ excitation, interpreted as $\ket{S_d}$ and $\ket{S_i}$. Smaller linewidths in the present experiment allowed us to resolve the doublet. We note that the $\ket{S_i}$, $\ket{T_0}$ doublet is not observed at every location on the sample, a consequence of inhomogeneities. There is no obvious correlation, doublet versus no doublet, with other optical properties, for instance the X$^{0}$ energy.

In conclusion, by identifing all the PL lines from gated monolayer MoS$_{2}$, we find that only spin-$\downarrow$ bands at each valley are occupied, signalling ferromagnetic order. At $n=1.5 \times 10^{12}$~cm$^{-2}$, we extract from the PL spectra the ferromagnetic exchange energy, $\Sigma \approx 11.2 \pm 1.4\,$meV, and the intervalley Coulomb exchange energy, $J \approx 5$~meV. Fast decay of $J$ at larger $n$ allows us to extract the trion size, $a_{\rm tr} \approx 3$~nm at $n = 3.5 \times 10^{12}$~cm$^{-2}$. The large exchange energy suggests that ferromagnetic ordering should survive up to tens of Kelvin. At these temperatures, the optical probe is no longer useful on account of phonon broadening of the optical lines -- this motivates an investigation of this state of matter with a sensitive magnetometer \cite{Mattiat2020,Marchiori2022}.

The work was supported by the QCQT PhD School of the University of Basel, the Georg H.\ Endress Foundation, and the Swiss Nanoscience Institute. N.L.\ acknowledges support from the Swiss National Science Foundation (Project No.\ P500PT\textunderscore 206917). K.W.\ and T.T.\ acknowledge support from the JSPS KAKENHI (Grant Numbers 21H05233 and 23H02052) and World Premier International Research Center Initiative (WPI), MEXT, Japan.

%


\end{document}


\title{SUPPLEMENTARY INFORMATION \\
\vspace{10PT}
Exchange energy of the ferromagnetic electronic ground-state in a monolayer semiconductor}

\author{Nadine Leisgang}
\affiliation{Department of Physics, University of Basel, Klingelbergstrasse 82, 4056 Basel, Switzerland}
\affiliation{Department of Physics, Harvard University, Cambridge, Massachusetts 02138, US}

\author{Dmitry Miserev}
\affiliation{Department of Physics, University of Basel, Klingelbergstrasse 82, 4056 Basel, Switzerland}

\author{Hinrich Mattiat}
\affiliation{Department of Physics, University of Basel, Klingelbergstrasse 82, 4056 Basel, Switzerland}

\author{Lukas Schneider}
\affiliation{Department of Physics, University of Basel, Klingelbergstrasse 82, 4056 Basel, Switzerland}

\author{Lukas Sponfeldner}
\affiliation{Department of Physics, University of Basel, Klingelbergstrasse 82, 4056 Basel, Switzerland}

\author{Kenji Watanabe}
\affiliation{Research Center for Electronic and Optical Materials, National Institute for Materials Science, 1-1 Namiki, Tsukuba 305-0044, Japan}

\author{Takashi Taniguchi}
\affiliation{Research Center for Materials Nanoarchitectonics, National Institute for Materials Science,  1-1 Namiki, Tsukuba 305-0044, Japan}

\author{Martino Poggio} 
\affiliation{Department of Physics, University of Basel, Klingelbergstrasse 82, 4056 Basel, Switzerland}

\author{Richard J. Warburton}
\affiliation{Department of Physics, University of Basel, Klingelbergstrasse 82, 4056 Basel, Switzerland}

\maketitle

\section{Building a van der Waals heterostructure device}
\subsection{Device fabrication}
\begin{figure}[b!]
\centering
\includegraphics[width=85mm]{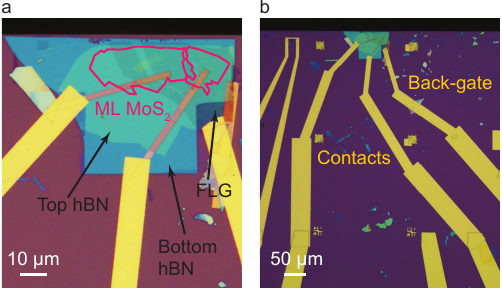} 
\caption{(a) and (b) are optical images of the device. Monolayer MoS$_2$ is encapsulated between two hBN flakes. Few-layer graphene (FLG) serves as back-gate electrode to inject charge carriers into the MoS$_2$ which is grounded via a bottom Ti/Au contact. The monolayer region is highlighted in red.}
\label{FigDeviceGatedMoS2}
\end{figure}
We fabricated our van der Waals heterostructure (vdWH) by stacking two-dimensional (2D) materials using a dry-transfer technique. For this, we attached a polydimethylsiloxane (PDMS) stamp with a thin polycarbonate (PC) film to a glass slide to pick up the individual layers from top to bottom. All flakes were previously mechanically exfoliated from bulk crystals (natural MoS$_2$ crystal from SPI Supplies, synthetic hexagonal boron nitride (hBN) \cite{taniguchi2007}, and natural graphite from Graphene Supermarket) on SiO$_2$ (285~nm)/Si substrates. The stacking sequence was divided into two steps. First, we placed the few-layer graphene (FLG) back-gate and the bottom hBN layer onto an undoped SiO$_2$ (285~nm)/Si substrate with pre-patterned alignment markers and big metal leads Ti(5~nm)/Au (55~nm). Bottom metal contacts were then patterned by electron-beam lithography and subsequent metal deposition of  Ti (10~nm)/Au (20~nm). In a second step, monolayer MoS$_2$ and a capping hBN layer were placed on top of the initial van der Waals stack. To improve the contact to the MoS$_2$, we annealed our device in vacuum at $T=100\,^{\circ}C$ for $\sim 12$~hours. Fig.~\ref{FigDeviceGatedMoS2} shows the optical image of the gated monolayer MoS$_2$ device. The monolayer region is highlighted with red color.

\subsection{Electrial gating}
We use a single-gated device to study the behavior of monolayer MoS$_2$ upon electrostatic doping. FLG serves as back-gate to inject electrons with density $n$. We can estimate the capacitance of our device using simple electrostatics: the monolayer MoS$_2$ and the FLG act as two electrodes separated by the bottom hBN layer with thickness d$_\tr{hBN}=34 \pm 1.7$~nm. The device capacitance per unit area is then estimated to be
\begin{equation}
C=\frac{\epsilon_0 \epsilon_\tr{hBN}}{d_\tr{hBN}}=97.9 \, \pm 4.9 \, \textrm{nFcm}^{-2} ,
\label{EqCapacitanceSingleGate}
\end{equation}
where $\epsilon_0$ is the vacuum permittivity and $\epsilon_\tr{hBN}=3.76$ \cite{laturia2018} is the dielectric constant of hBN. Here, we accounted for a 5\% uncertainty in determining the layer thickness via atomic force microscopy (AFM). By applying a gate voltage, $V_\tr{G}$, to the backgate, while keeping the monolayer MoS$_2$ grounded, we can inject charge carriers into our device through
\begin{equation}
n=C \cdot V_\tr{G} \, .
\end{equation}
In general, we can see from our optical measurement that $n \simeq 0$ at $V_\tr{G}=0$ in the device: the spectrum is dominated by the neutral exciton, labelled as $X^0$.  

\section{Experimental set-up}
\begin{figure}[t!]
\centering
\includegraphics[width=85mm]{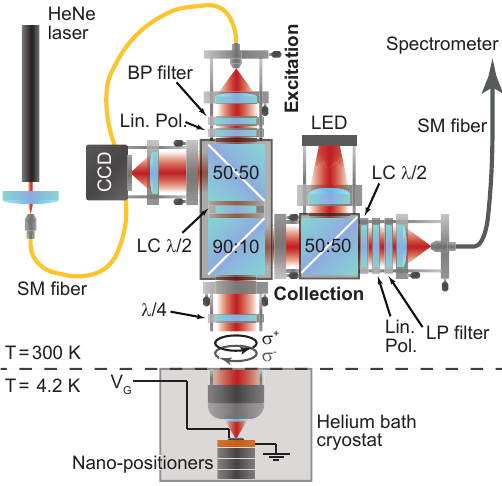} 
\caption{Schematic of the experimental set-up for detecting the photoluminescence (PL) signal at cryogenic temperatures ($T=4$~K). SM stands for single-mode, BP and LP for band- and long-pass, LC for liquid crystal.}
\label{expSetup}
\end{figure}
The photoluminescence (PL) spectra were recorded at cryogenic temperatures ($T=4$~K) with the set-up sketched in Supplementary Fig.~\ref{expSetup}. We use a red helium-neon (HeNe) laser ($\lambda=633$~nm, $E=1.96$~eV) to optically excite our sample. The light from the laser is coupled into a single-mode (SM) fiber, with its output connected to a home-built confocal microscope. To produce spectrally clean laser light, the incoming light is first sent through a narrow band-pass (BP) filter. The light then passes through a linear polarizer (LP) and a computer-controlled liquid crystal (LC) to generate two perpendicular linear polarizations on demand. In combination with an achromatic quarter-wave plate ($\lambda/4$), we can create circularly polarized light with either right-handed ($\sigma^+$) or left-handed ($\sigma^-$) orientation by applying a voltage to the LC. The light is focused onto onto the sample using a high-NA microscope objective, NA~$=0.85$, with an excitation power below $500$~nW. The diffraction-limited spot can be scanned over the sample by cryogenic nano-positionners. The emitted light from the sample is collected by the same objective, goes again through the $\lambda/4$-waveplate, and is directed to the collection arm using a 90:10 beam-splitter (BS). The combination of another LC with a linear polarizer in the collection arm is used to resolve the polarization of the PL signal. A long-pass (LP) filter removes the background light before the signal is coupled into a SM fiber and sent to the spectrometer with a 1500 grooves-per-millimeter grating. The spectra are recorded by a liquid-nitrogen cooled charged coupled device (CCD) array. 

\section{Experimental analysis}
\subsection{Position-dependent photoluminescence data and reproducibility}
\begin{figure*}[t!]
\centering
\includegraphics[width=180mm]{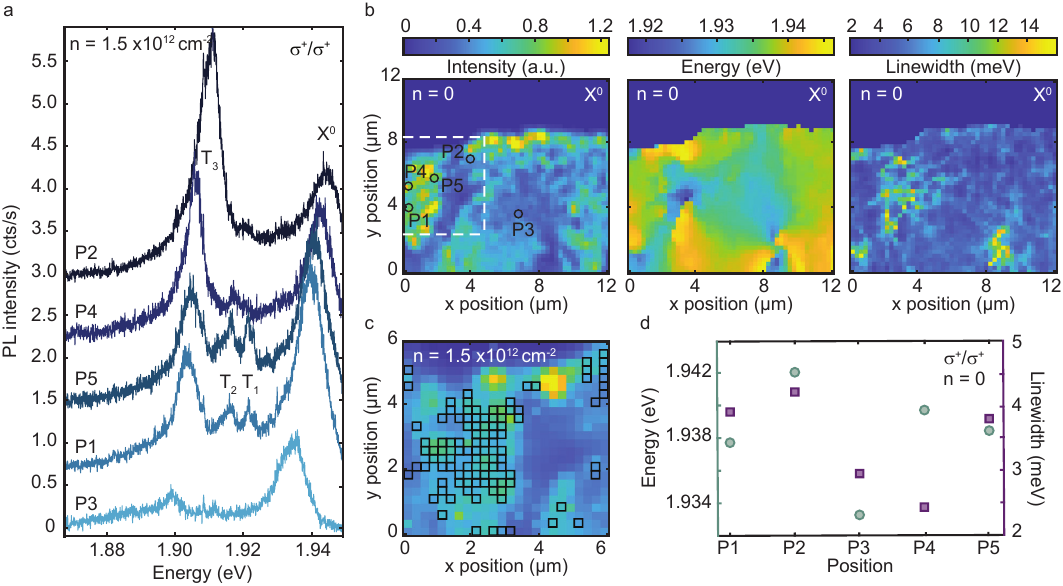} 
\caption{(a) PL spectra at $n=1.5 \times 10^{12}$~cm$^{-2}$ (at $+$9.00~T and 4.2~K) for $\sigma^{+}/\sigma^{+}$-excitation/collection at different positions. (b) Integrated PL intensity, energy and linewidth of the neutral exciton, $X^0$, as a function of position. (c) Appearance of a ``doublet" at $n=1.5 \times 10^{12}$~cm$^{-2}$ as a function of position (black squares). (d) Energy and linewidth of $X^0$ at different positions as indicated in (b).}
\label{fig2Dmap}
\end{figure*}

Different sample positions, labelled as P1, \dots P5, are investigated, marked by black circles on the position-dependent PL map, Fig.~\ref{fig2Dmap}(a),(b). All positions represent clean spots on the sample, reflected by high intensities and narrow optical linewidths, $\Gamma = 2.8-4.1 $~meV, of the neutral exciton, $X^0$, at $n=0$. 
At $n=1.5 \times 10^{12}$~cm$^{-2}$, a ``doublet" feature T1, T2 is observable in $\sigma^+/\sigma^+$-polarization at various locations on the sample, Fig.~\ref{fig2Dmap}(a),(c). We note that the T1, T2 doublet is not observed at every location on the sample. There is no obvious correlation, doublet versus no doublet, with other optical properties, such as X$^{0}$ energy and linewidth, Fig.\ref{fig2Dmap}d.

Fig.~\ref{figPositions} presents the density-dependent PL data in a perpendicular magnetic field of $B_z=+9.00$~T for two different positions on the sample, P1 and P2. The optical response is plotted as a matrix: $\sigma^{+}/\sigma^{-}$ refers to excitation with $\sigma^{+}$, collection with $\sigma^{-}$; and etc. We note that the $X^0$ energies, linewidths and intensities are comparable to the data shown in the main text (position P5), Fig.~\ref{fig2Dmap}(d). At both positions, electron doping is demonstrated. With increasing carrier concentration, the intensity of the neutral exciton is transferred to the negatively charged trions. For P1, three emission lines,  T$_1$, T$_2$ and T$_3$, are observable in $\sigma^+/\sigma^+$-polarization, while for P2 the ``doublet" feature is not resolvable. We note that the different trion states, T$_1$, T$_2$ and T$_3$, are very close in energy. A clear observation of the doublet state, Fig.~\ref{fig2Dmap} and Fig.~\ref{figPositions}, for position P1 compared to position P2, is probably related to the sample quality and the specific dielectric environment.

\begin{figure}[hb!]
\centering
\includegraphics[width=85mm]{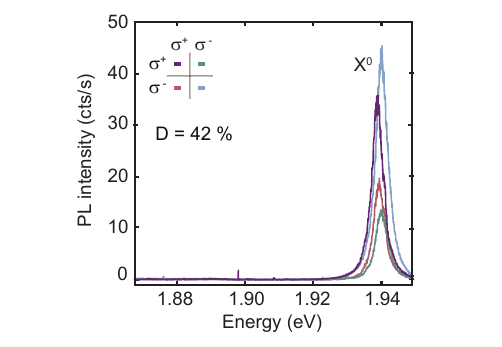}
\caption{PL spectra at $n=0$ in all four circular polarizations. Lorentzian fits to the neutral exciton, $X^0$, reveal a valley polarization of $D=42 \%$ in our gated monolayer MoS$_2$ device.}
\label{figSpectraX0}
\end{figure}

\begin{figure}[b!]
\centering
\includegraphics[width=85mm]{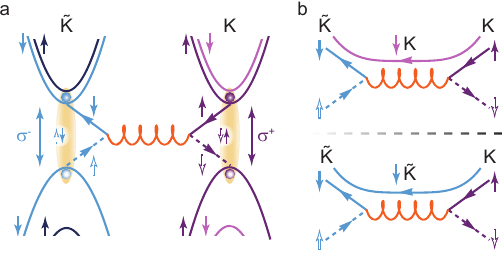} 
\caption{Electron-hole exchange diagrams (a) without and (b) with an excess electron from the Fermi sea. An exciton at the $K$-point scatters into an exciton at the $\tilde{K}$-point via the annihilation channel, flipping the spins of both electron and hole. The conduction electron and valence hole wave functions are shown by solid and dashed lines, respectively. The orange wavy lines correspond to the Coulomb interaction.}
\label{ehExchange}
\end{figure}

\begin{figure*}[t!]
\centering
\includegraphics[width=153mm]{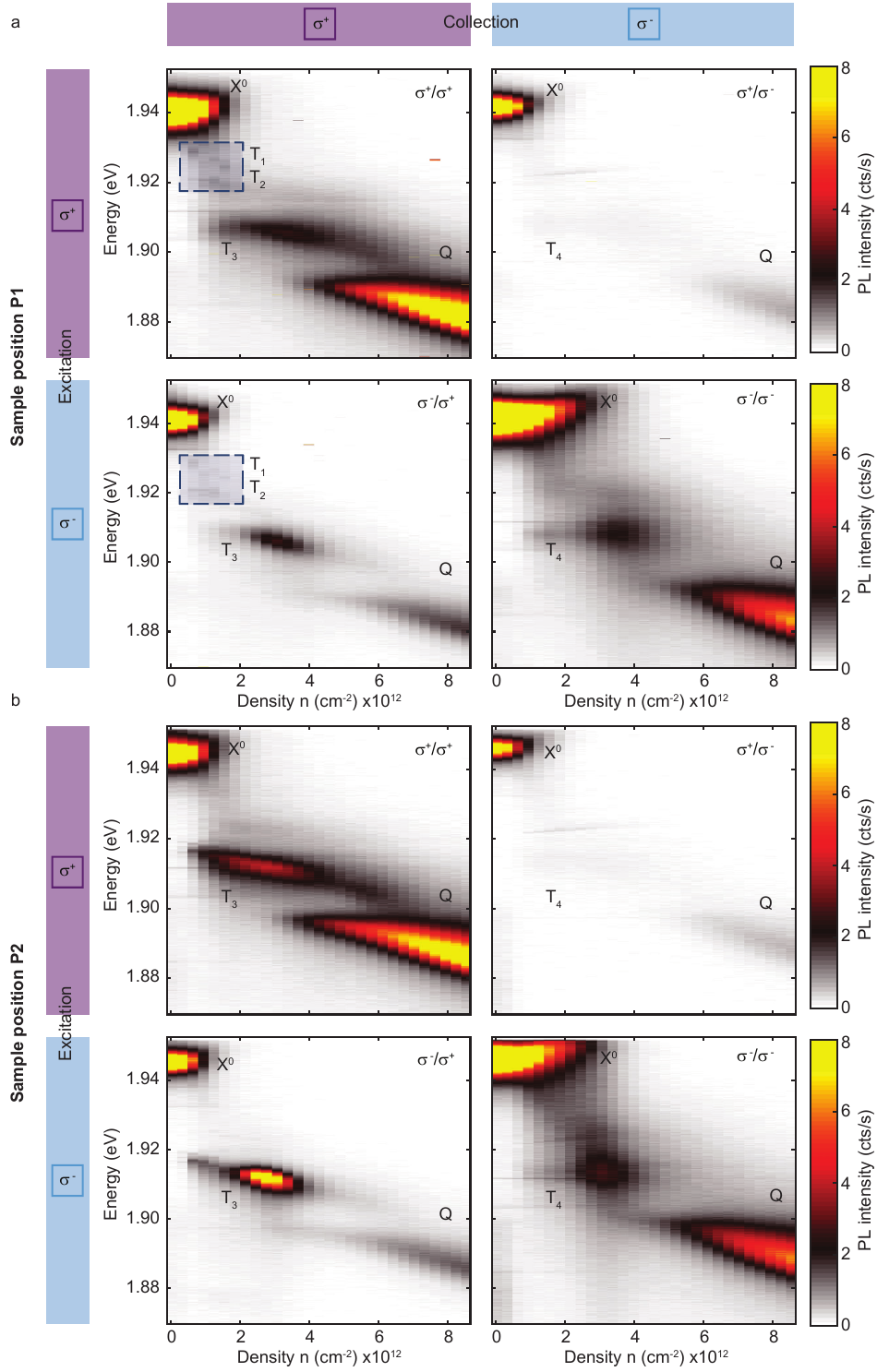}
\caption{PL for quasi-resonant excitation on positions (a) P1 and (b) P2 at $+$9.00~T and 4.2~K shown as a matrix: $\sigma^{+}$/$\sigma^{-}$-excitation, $\sigma^{+}$/$\sigma^{-}$-collection.}
\label{figPositions}
\end{figure*}

\subsection{Valley polarization}
PL arises from the radiative recombination of bright excitons at the $K$- and $\tilde{K}$-points. The optical selection rules ensure that normally incident $\sigma^+$- polarized light can only create an exciton at the $K$-point, $\sigma^-$- polarized light at the $\tilde{K}$-point. Conversely, one can measure the valley polarization of the recombining exciton because $\sigma^+$ photons originate from recombination of excitons at the $K$-point, $\sigma^-$ photons from recombination at the $\tilde{K}$-point.

In our experiment, the valley polarization of the optically excited electrons can be estimated from the PL emission in the circular polarization of the same and opposite helicity compared to the excitation light, a quantity called optical dichroism, $D$,
\begin{eqnarray}
D=\frac{I(\sigma^{+})-I(\sigma^{-})}{I(\sigma^{+})+I(\sigma^{-})} \, . 
\end{eqnarray}

At $n=0$, $D$ is 42\% for $\sigma^+$- and $\sigma^-$-polarization, Fig.\ref{figSpectraX0}. This number indicates a significant bright exciton transfer between the $K$- and $\tilde{K}$-valley. This value increases to 64\% at $n=1.5 \times 10^{12}$~cm$^{-2}$ for both $\sigma^+$- and $\sigma^-$-polarization, respectively. We ascribe this observation to a large Bir-Aronov-Pikus electron-hole exchange rate.

Assuming that electron-hole exchange is the only significant mechanism, and applying a rate equation model, we find the electron-hole exchange relaxation time $\tau_{eh}$
\begin{eqnarray}
\frac{\tau_{eh}}{\tau_{ex}} \simeq \frac{2 D}{1 - D} \, .
\end{eqnarray}
Taking a typical value for the radiative lifetime, $\tau_{ex} \simeq 4$~ps \cite{cadiz2016,korn2011}, the measured value of $D$ allows the electron-hole exchange time to be estimated as $\tau_{eh} \simeq 6$~ps. This is similar to other estimates of the exciton-valley relaxation time \cite{mai2014}. The electron-hole exchange is caused by the long-range Coulomb interaction \cite{yu2014,glazov2014,HongyiYu2015}, Fig.~\ref{ehExchange}.

\section{Theoretical description of the trion states}
\subsection{Microscopic model for low electron densities}
We use the few-particle trion picture to interpret our experimental results. This approach is valid as long as the Fermi wavelength, $2 \pi/k_F$, $k_F$ is the Fermi momentum, is much larger than the trion size, $a_{\rm tr} \sim 2\,$nm  \cite{zhang2014,christopher2017,rana2020}. This is the case for the lowest electron densities in our experiment.

In monolayer MoS$_2$, electrons have two degrees of freedom, spin, $s_z= \pm \frac{1}{2}$, and valley isospin, $\tau_z= \pm \frac{1}{2}$. Possible trion quantum states follow from the eigenvalue problem of the three-particle Hamiltonian. According to Fermi statistics, the total electron wave function within a trion state must be antisymmetric under exchange of electrons.

Within the trion state, each electron (1 or 2) can either have spin $|\uparrow\rangle$ or spin $|\downarrow\rangle$. Consequently, there are four product states,
\begin{eqnarray}
|\uparrow\rangle_1 | \uparrow\rangle_2, \hspace{5pt}
|\uparrow\rangle_1 | \downarrow\rangle_2, \hspace{5pt}
|\downarrow\rangle_1 | \uparrow\rangle_2, \hspace{5pt}
|\downarrow\rangle_1 | \downarrow\rangle_2 \, .
\end{eqnarray}
However, it is more convenient to work with the eigenstates of the total spin: the spin-singlet state, $S = 0$,
\begin{eqnarray}
\frac{1}{\sqrt{2}}[|\uparrow\rangle_1 |\downarrow \rangle_2 - |\downarrow \rangle_1 |\uparrow \rangle_2]
\end{eqnarray}
and the three spin-triplet states, $S = 1$,
\begin{eqnarray}
& \chi_+ = |\uparrow\rangle_1 |\uparrow\rangle_2, \hspace{5pt} \chi_- = |\downarrow\rangle_1 |\downarrow\rangle_2 , \label{chipm}\\
& \chi_0 = \frac{1}{\sqrt{2}}\left[|\uparrow\rangle_1 |\downarrow \rangle_2 + |\downarrow \rangle_1 |\uparrow \rangle_2\right] \, . \label{chi0}
\end{eqnarray}
We note that the singlet state is antisymmetric, while the triplet states are symmetric. 
Analogous to the spin states, there are four possible orbital states, one antisymmetric valley-singlet state, $\tau = 0$,
\begin{eqnarray}
\frac{1}{\sqrt{2}}\left[|K\rangle_1 |\tilde{K} \rangle_2 - |\tilde{K} \rangle_1 |K \rangle_2\right]
\end{eqnarray}
and three symmetric valley-triplet states, $\tau = 1$,
\begin{eqnarray}
\hspace{-5pt} |K\rangle_1 |K\rangle_2, \hspace{5pt}
\frac{1}{\sqrt{2}}\left[|K\rangle_1 |\tilde{K} \rangle_2 + |\tilde{K} \rangle_1 |K \rangle_2\right], \hspace{5pt}
|\tilde{K}\rangle_1 |\tilde{K}\rangle_2 \, .
\end{eqnarray}

Antisymmetric trion wave functions can only be constructed as a product of singlet and triplet states. Consequently, three spin-singlet states are attached to each valley triplet state,
\begin{eqnarray}
|S_d \rangle & = & |K\rangle_1 |K\rangle_2 \cdot \frac{1}{\sqrt{2}}[|\uparrow\rangle_1 |\downarrow \rangle_2 - |\downarrow \rangle_1 |\uparrow \rangle_2] \, , \\
|S_i\rangle & = & \frac{1}{\sqrt{2}}[|K\rangle_1 |\tilde{K}\rangle_2 + |\tilde{K}\rangle_1 |K\rangle_2] \nonumber \\
&& \cdot \frac{1}{\sqrt{2}}[|\uparrow\rangle_1 |\downarrow \rangle_2 - |\downarrow \rangle_1 |\uparrow \rangle_2] \, , \label{Sd}\\
|\tilde{S}_d \rangle & = & |\tilde{K}\rangle_1 |\tilde{K}\rangle_2 \cdot \frac{1}{\sqrt{2}}[|\uparrow\rangle_1 |\downarrow \rangle_2 - |\downarrow \rangle_1 |\uparrow \rangle_2] \, ,
\end{eqnarray}
and three spin-triplet states are attached to the valley singlet state,
\begin{eqnarray}
|T_s\rangle & = & \frac{1}{\sqrt{2}}[|K\rangle_1 |\tilde{K}\rangle_2 - |\tilde{K}\rangle_1 |K\rangle_2] \cdot \chi_s \, . \label{T}
\end{eqnarray}
Here, $\chi_\pm$ and $\chi_0$ are the spin-triplet states defined in Eqs.~(\ref{chipm}), (\ref{chi0}), $|S_d\rangle$ and $|\tilde{S}_d\rangle$ correspond to the intravalley spin-singlet states at the $K$- and $\tilde{K}$-point, respectively, $|S_i\rangle$ is the intervalley spin-singlet state.

\subsection{Optical excitation of the trion states under circular polarization}
The optical selection rules ensure that normally incident $\sigma^+$-polarized light can only create an exciton at the $K$-point. The electron component of the bright exciton can be written as
\begin{eqnarray}
|ex_+\rangle = |K \uparrow\rangle \, .
\end{eqnarray}
In general, such an exciton can bind to an electron from the following Fermi seas:
\begin{eqnarray}
|K \downarrow \rangle , \hspace{5pt} |\tilde{K} \uparrow\rangle, \hspace{5pt} |\tilde{K} \downarrow\rangle \, . 
\end{eqnarray}
We note that binding to an electron with the same spin and valley quantum number, in this case $|K \uparrow\rangle$, is not possible due to the Pauli exclusion principle.
This results in the formation of the following three trion species
\begin{eqnarray}
&& \hspace{-20pt} \frac{1}{\sqrt{2}}\left[|K \uparrow\rangle_1 |K \downarrow\rangle_2 - |K \downarrow\rangle_1 |K \uparrow\rangle_2 \right] = |S_d \rangle \, , \\
&& \hspace{-20pt} \frac{1}{\sqrt{2}}\left[|K \uparrow\rangle_1 |\tilde{K} \uparrow\rangle_2 - |\tilde{K} \uparrow\rangle_1  |K \uparrow\rangle_2 \right] = |T_+\rangle \, ,\\
&& \hspace{-20pt} \frac{1}{\sqrt{2}}\left[|K \uparrow\rangle_1 |\tilde{K} \downarrow\rangle_2 - |\tilde{K} \downarrow\rangle_1  |K \uparrow\rangle_2 \right] \nonumber \\
&& \hspace{90pt} = \frac{1}{\sqrt{2}}\left[ |T_0\rangle + |S_i\rangle \right] \, \label{exc_sigma+_doublet}.
\end{eqnarray}
We note that the optically excited trion state in Eq.~(\ref{exc_sigma+_doublet}) is a superposition of the two intervalley trion states, $|T_0\rangle$ and $|S_i\rangle$. This state appears as a ``doublet'' feature in our experiment due to the intervalley exchange Coulomb interaction splitting $|T_0\rangle$ and $|S_i\rangle$.

Accordingly, excitation with $\sigma^-$-polarized light creates an exciton at the $\tilde{K}$-point
\begin{eqnarray}
|ex_-\rangle = |\tilde{K} \downarrow\rangle \, ,
\end{eqnarray}
which can bind to an electron $|\tilde{K} \uparrow\rangle$, $|K \downarrow\rangle$, $|K \uparrow\rangle$ and form the trions $|\tilde{S}_d\rangle$, $|T_-\rangle$, $-[|T_0\rangle + |S_i\rangle]/\sqrt{2}$, respectively.

\subsection{Energies of the trion states}
We calculate the relevant energies of the trion states in our experiment using first-order perturbation theory. The Hamiltonian of the system is given by 
\begin{eqnarray}
&& \hat{H} = \hat{H}_0 + \hat{V} \, , \\
&& \hat{V} = \Sigma \left( s_{z, 1} + s_{z, 2} \right) + \hat{H}_\tr{so} + \hat{H}_z + \hat{H}_{\rm ex} \, , \label{V}
\end{eqnarray}
with $\hat{H}\psi = E \psi$, $E$ is the trion energy. Here, $\hat{H}_0$ is the three-body Hamiltonian that conserves both total spin and total valley isospin. It includes the kinetic energy term as well as the spin-conserving and valley-conserving Coulomb matrix elements.
If only $\hat{H}_0$ is taken into account, all trion species have the same energy.
The perturbation, $\hat{V}$, breaks both spin and valley isospin conservation, yet the $z$-components of spin and valley isospin remain conserved. This results in the observed fine structure of the trion peaks.
$\Sigma > 0$ in Eq.~(\ref{V}) stands for the ferromagnetic exchange energy splitting off the $\uparrow$ bands, $s_{z,1}$ ($s_{z,2}$) is the $z$-projection of the spin operator, corresponding to the first (second) electron within the trion state.
The spin-orbit interaction, $\hat{H}_\tr{so}$, is represented by the spin-valley coupling
\begin{eqnarray}
\hat{H}_\tr{so} = 
-2 \Delta_\tr{so} \left( s_{z,1}\tau_{z, 1} + s_{z,2} \tau_{z, 2} \right)  \, ,
\end{eqnarray}
with the following non-zero matrix elements:
\begin{eqnarray}
\bra{K \uparrow} \hat{H}_\tr{so} \ket{K \uparrow}& =\bra{\tilde{K} \downarrow} \hat{H}_\tr{so} \ket{\tilde{K}\downarrow}& =-\frac{\Delta_\tr{so}}{2} \, ,\\
\bra{K \downarrow} \hat{H}_\tr{so} \ket{K \downarrow}& =\bra{\tilde{K} \uparrow} \hat{H}_\tr{so} \ket{\tilde{K}\uparrow}& =+\frac{\Delta_\tr{so}}{2} \, .
\end{eqnarray}
Here, $\Delta_\tr{so}$ is the conduction band spin-orbit splitting, $\tau_{z,1} = \pm 1/2$, $\tau_{z,2} = \pm 1/2$.
The Zeeman term corresponding to the out-of-plane magnetic field, $B_z$, is given by 
\begin{eqnarray}
\hat{H}_{z} = g\mu_B B_z \left( s_{z,1}+s_{z,2} \right) + 2 \alpha \mu_B B_z \left( \tau_{z, 1} + \tau_{z, 2} \right) \, ,
\end{eqnarray}
where the first (second) part is related to the spin (valley isospin) contribution, with $g=1.98$ and $\alpha=0.375$ \cite{kormanyos2014}, $\mu_B$ is the Bohr magneton. 

The Coulomb interaction conserves total spin, i.e., we can work within a spin-singlet and spin-triplet channels separately. The only component of the Coulomb interaction that splits different valley states is the intervalley exchange Coulomb matrix element
\begin{eqnarray}
\frac{J}{2} \equiv \langle \tau_z, -\tau_z| \hat{H}_\tr{ex}|-\tau_z, \tau_z\rangle \, ,
\end{eqnarray}
where $\tau_z = \pm 1/2$ labels the valley. The spin index is lifted due to the spin-conserving nature of the Coulomb interaction.
The intervalley exchange Coulomb interaction splits the states $|S_d\rangle$, $|\tilde{S}_d\rangle$, $|S_i\rangle$ and $|T_s\rangle$ as follows
\begin{eqnarray}
    && \hspace{-15pt} \langle S_d|\hat{H}_\tr{ex}| S_d\rangle = \langle \tilde{S}_d |\hat{H}_\tr{ex} | \tilde{S}_d \rangle = \langle T_\pm |\hat{H}_\tr{ex} | T_\pm \rangle = 0 \, , \label{0} \\
    && \hspace{-15pt} \langle S_i |\hat{H}_\tr{ex}| S_i \rangle =  \frac{J}{2} , \hspace{5pt} \langle T_0 | \hat{H}_\tr{ex} | T_0 \rangle = - \frac{J}{2} \, . \label{J}
\end{eqnarray}
Combining this with the single-particle contributions and the ferromagnetic exchange, $\Sigma$, we find that states $|S_d\rangle$, $|\tilde{S}_d\rangle$, $|T_\pm\rangle$ are the eigenstates of $\hat{H}$ with the following energies (counted from the binding energy provided by $\hat{H}_0$)
\begin{eqnarray}
    && E_{S_d} = 2 \alpha \mu_B B_z \, , \label{Ed}\\
    && E_{\tilde{S}_d} = - 2 \alpha \mu_B B_z \, , \label{Edtilde} \\
    && E_{T_\pm} = -\frac{J}{2} \pm \left( \Sigma + g \mu_B B_z \right) \, . \label{Epm}
\end{eqnarray}
The states $|S_i\rangle$ and $|T_0\rangle$ are mixed by the spin-orbit interaction
\begin{eqnarray}
    && \hat{V} |S_i \rangle = \frac{J}{2} |S_i\rangle - \Delta_\tr{so} |T_0\rangle \, , \\
    && \hat{V} |T_0 \rangle = - \frac{J}{2} |T_0\rangle - \Delta_\tr{so} |S_i\rangle \, .
\end{eqnarray}
The energies of the true eigenstates -- we call them $|S_i\rangle_\tr{so}$ and $|T_0\rangle_\tr{so}$ -- are then
\begin{eqnarray}
    && E_{S_i} = \sqrt{\frac{J^2}{4} + \Delta_\tr{so}^2} , \hspace{5pt} E_{T_0} = - \sqrt{\frac{J^2}{4} + \Delta_\tr{so}^2} \, .
\end{eqnarray}
$\Delta_\tr{so}$ is hardly dependent on the electron densities, $n$, within the experimental gate-voltage range. However, we observe that at larger electron densities, $n \approx 3.5 \times 10^{12}\,$cm$^{-2}$, the splitting $E_{S_i} - E_{T_0}$ vanishes (the two peaks merge together). This implies that $\Delta_\tr{so}$ is negligible compared to $J$ for the relevant electron densities in our experiment. We can approximate
\begin{eqnarray}
    E_{S_i} \approx \frac{J}{2}, \hspace{5pt} E_{T_0} \approx - \frac{J}{2} \, . \label{Edoublet}
\end{eqnarray}

In our experiment, there are only four relevant trions, $|S_d\rangle$, $|T_-\rangle$ and the doublet state $[|T_0\rangle + |S_i\rangle]/\sqrt{2}$, providing the $E_{S_i}$ and $E_{T_0}$ energies. The ferromagnetic exchange energy, $\Sigma$, and the intervalley exchange Coulomb energy, $J$, are then found from the trion splittings
\begin{eqnarray}
   && J \approx E_{S_i} - E_{T_0} \, , \\
   && \Sigma =  E_{T_0} - E_{T_-} - g \mu_B B_z \, .
\end{eqnarray}

%
